\documentclass[aps,showkeys, showpacs,amsmath,amssymb,amsfonts]{revtex4}

\usepackage{enumerate}
\usepackage{graphicx}
\usepackage{dcolumn}
\usepackage[usenames]{color}
\usepackage{ulem} 

\def\p{\mathbf{p}}

\begin{document}
\title{Group theory, entropy and the third law of thermodynamics}
\author{$^{1}$Bilal Canturk}
\email{ bilalcntrk@gmail.com }
\author{$^{2}$Thomas Oikonomou}
\email{thomas.oikonomou@nu.edu.kz}
\author{$^{1}$G. Baris Bagci}
\email{gbb0002@hotmail.com}
\affiliation{$^{1}$Department of Materials Science and Nanotechnology Engineering,
TOBB University of Economics and Technology, 06560 Ankara,
Turkey}
\affiliation{$^{2}$Department of Physics, School of Science and Technology, Nazarbayev University, 53 Kabanbay Batyr Avenue, Astana 010000, Kazakhstan}

\begin{abstract}

Curado \textit{et al.} [Ann. Phys. \textbf{366} (2016) 22] have recently studied the axiomatic structure and the universality of a three-parameter trace-form entropy inspired by the group-theoretical structure. In this work, we study the group-theoretical entropy $S_{a,b,r}$ in the context of the third law of thermodynamics where the parameters $\left\lbrace a,b,r \right\rbrace $ are all independent. We show that this three-parameter entropy expression can simultaneously satisfy the third law of thermodynamics and the three Khinchin axioms, namely continuity, concavity and expansibility only when the parameter $b$ is set to zero. In other words, it is thermodynamically valid only as a two-parameter generalization $S_{a,r}$. Moreover, the restriction set by the third law i.e., the condition $b = 0$, is important in the sense that the so obtained two-parameter group-theoretical entropy becomes extensive only when this condition is met. We also illustrate the interval of validity of the third law using the one-dimensional Ising model with no external field. Finally, we show that the $S_{a,r}$ is in the same universality class as that of the Kaniadakis entropy for $0 < r < 1$ while it has a distinct universality class in the interval $-1 < r < 0$.

\end{abstract}




\newpage \setcounter{page}{1}

\maketitle

\section{introduction}

Recently, there has been a great deal of progress in constructing generalized statistical mechanics through the applications of the information-theoretic entropies. To list some of them, one can cite Tsallis \cite{Tsallis}, R{\'e}nyi \cite{Renyi}, Kaniadakis \cite{Kaniadakis} entropies. These generalized entropies aim to explain the non-equilibrium stationary metastable states through the deformation in the underlying entropic structure. Along this direction, many important applications were reported in the fields of generalized reaction rates \cite{rate1,rate2,rate3,rate4}, quantum information \cite {Rajagopal,Chang,Tribeche,Sever,Rastegin,Kak,Portesi}, plasma physics \cite{Dedu,Rotundo,Tri}, high energy physics \cite{Van1,high2,high3} and the rigid rotators in modelling the molecular structure \cite{reis,rotator}. The common feature of these entropies is to yield inverse power law distributions through the entropy maximization \cite{Tsallis, Kaniadakis,Bagci,Bashkirov}.

It is also worth noting the recent progress in the use of the fractional calculus based entropies such as Ubriaco \cite{Ubriaco} and Machado \cite{Machado} entropies. These fractional entropies have been applied to the study of the financial time series and stock market index \cite{Machado, Machado1}, used for explaining the relation between DNA and the fractional Brownian motion \cite{Machado2}, image splicing \cite{ibrahim}, and a new definition of the complexity metrics \cite{Tang}.

A very general novel approach, first initiated in Ref. \cite{tempesta0}, is to obtain generalized entropy expressions through the underlying group-theoretical structure \cite{tempesta1, tempesta2, tempesta3}. The general entropy expression of this universal group character includes many known trace-form entropy measures such as Boltzmann-Gibbs, Tsallis \cite{Tsallis}, Kaniadakis \cite{Kaniadakis} and Borges-Roditi \cite{Roditi} entropies. Moreover, the group-theoretical entropy satisfies the three Khinchin axioms i.e. continuity, concavity and expansibility \cite{tempesta3}.

On the other hand, the question remains whether the group-theoretical entropy can be used for a consistent thermodynamics. One possible check in this context has been offered in Ref. \cite{Bento}, namely, the third law of thermodynamics. The third law of thermodynamics should be valid for any novel entropy measure independent of the Hamiltonian under scrutiny. For this purpose, Bento \textit{et al.} \cite{Bento} assume that the microscopic energies $E_{\lambda}$ are ordered  where $\lambda = 0,1,...,N$ (no degeneracy present). The ground state is the one corresponding to $\lambda = 0$, and $p_{\lambda}$ denotes the probability of the system to be in the state $\lambda$. The normalization is also assumed i.e $\sum_{\lambda} p_{\lambda}=1$. Denoting the enumeration as $n=1,...,N$, the normalization implies that any function $f$ of the probability satisfies the relation $\frac{\partial f\left( p_{0}\right) }{\partial p_{n}} = -\frac{\partial f\left( p_{0}\right) }{\partial p_{0}}$. The third law of thermodynamics is then checked by assigning unity to the ground state probability $p_0$ and zero to the probabilities relevant to all the remaining states. The inverse temperature is given by $\beta = \sum_{n} \beta_{n}$ where $\beta_{n}$ reads \cite{Bento} 
\begin{eqnarray}\label{temp}
\beta_{n}= \frac{\partial S}  {\partial p_{n}} \left({\frac{\partial U}{\partial p_{n}}}\right)^{-1}.
\end{eqnarray}
If this inverse temperature attains (plus) infinity in the limit $\left\lbrace p_{0}, p_{n}\right\rbrace \to \left\lbrace 1, 0 \right\rbrace$, one confirms the validity of the third law of thermodynamics concerning the particular entropy measure (see Ref. \cite{Bento} for the details of the procedure). We also note that the conformation to the third law of thermodynamics can even be checked for non-trace-form entropies such as R{\'e}nyi entropy \cite{Renyi, Renyi2} although we here limit ourselves to the trace-form group-theoretical entropies only \cite{tempesta1, tempesta2, tempesta3}.

In the next section, we show that the group-theoretical entropy proposed in Ref. \cite{tempesta3} simultaneously satisfies the third law of thermodynamics and the three axioms of Khinchin i.e. continuity, concavity and expansibility only as a two-parameter entropy $S_{a,r}$ with $b = 0$. We also illustrate the interval of validity for the third law of thermodynamics through the one-dimensional  Ising model with no external field. In addition, we show that  $S_{a,r}$ for $0 < r < 1$  is in the same universality class as that of the Kaniadakis entropy whereas $S_{a,r}$ has a novel universality class in the interval $-1 < r < 0$. The conclusions are provided in the final section.

\section{Group-theoretical entropy and the third law of thermodynamics}

The entropy definition based on the group-theoretical structure for $r>0$ reads
\begin{equation}\label{Plus-Group Entropy}
S_{a,b,r}^+[p]=\sum_{\lambda}
\frac{2p_\lambda(1-p_\lambda^r)}{\gamma(1+p_\lambda^r)-a(1-p_\lambda^r)}
\end{equation}
where we define $\gamma=\sqrt{a^2+4b}\geqslant0$ and $\{a,b,r \}$ are all real parameters \cite{tempesta3}. The group-theoretical entropy can also be written in terms of the deformed logarithms although we have here provided its explicit form \cite{tempesta3, thomas, borges}. This entropy definition satisfies the first three Khinchin axioms i.e., continuity, concavity and expansibility. In particular, the concavity holds if the following conditions are satisfied for $r > 0$ \cite{tempesta3}:
\begin{subequations}\label{plus-condition1}
\begin{eqnarray}
(i)\quad 0<r\leqslant1\quad  \rightarrow\quad && a<0\quad and\quad b>\dfrac{-a^2}{4} \\   && a\geqslant 0 \quad and\quad b>0 \\
(ii)\quad\quad r> 1\quad  \rightarrow\quad  && a\leqslant 0\quad and\quad b>\dfrac{-a^2}{4} \\ 
&&  a>0\quad and\quad b>\dfrac{a^{2}(r^{2}-1)}{4}. 
\end{eqnarray}
\end{subequations}
However, for the interval $r<0$ \cite{tempesta3}, the group-theoretical entropy reads
\begin{equation}\label{Minus-Group Entropy}
S_{a,b,r}^-[p]=\sum_{\lambda}
\frac{2p_\lambda(1-p_\lambda^{-r})}{\gamma(1+p_\lambda^{-r})-a(1-p_\lambda^{-r})}.
\end{equation}
Note that the superscripts $+$ and $-$ denote the group-theoretical entropy for $r > 0$ and $r <0$, respectively. The group entropy for $r < 0$ also satisfies the first three Khinchin axioms while the requirement of concavity now reads \cite{tempesta3}:
\begin{subequations}\label{minus-condition1}
\begin{eqnarray}
(i)-1\leqslant r< 0 \quad \rightarrow\quad && a<0\quad and\quad b>\dfrac{-a^2}{4}\\
 && a\geqslant 0\quad and\quad b>0\\
(ii) \quad\quad r<-1 \quad \rightarrow\quad && a\leqslant 0\quad and\quad b>\dfrac{-a^2}{4}\\
&& a>0 \quad and\quad b>\dfrac{a^{2}(r^{2}-1)}{4}. 
\end{eqnarray}
\end{subequations}
In order to check whether the group-theoretical entropy satisfies the third law, we must calculate Eq. \eqref{temp}. To this aim, we calculate the partial derivative of the group-theoretical entropy for $r > 0$ using Eq. \eqref{Plus-Group Entropy} 
\begin{eqnarray}\label{frac1}
\frac{\partial S_{a,b,r}^{+}}{\partial p_{n}}=\frac{2[-(\gamma+a)p_n^{2r}-2(\gamma r-a)p_n^{r}+(\gamma-a)]}{[(\gamma-a)+(\gamma+a)p_n^{r}]^2}+\frac{2[(\gamma+a)p_0^{2r}+2(\gamma r-a)p_0^{r}-(\gamma-a)]}{[(\gamma-a)+(\gamma+a)p_0^{r}]^2}
\end{eqnarray}
while the same quantity for $r < 0$ can be calculated through Eq. \eqref{Minus-Group Entropy} 
\begin{eqnarray}\label{frac2}
\frac{\partial S_{a,b,r}^{-}}{\partial p_{n}}=
\frac{2[-(\gamma+a)p_n^{-2r}+2(\gamma r+a)p_n^{-r}+(\gamma-a)]}{[(\gamma-a)+(a+\gamma)p_n^{-r}]^2} + \frac{2[(a+\gamma)p_0^{-2r}-2(\gamma r+a)p_0^{-r}-(\gamma-a)]}{[(\gamma-a)+(\gamma+a)p_0^{-r}]^2}.
\end{eqnarray}
In accordance with the group-theoretical structure (see Eq. (46) in Ref. \cite{tempesta2}), we define the mean energy through $U = \sum_{\lambda} p_{\lambda} E_{\lambda}$ so that
\begin{equation}\label{energy}
\frac{\partial U}{\partial p_{n}}=E_{n}-E_{0}.
\end{equation} 
Using Eqs. \eqref{frac1}, \eqref{frac2} and \eqref{energy}, one can then calculate $\beta_{n}$ in Eq. \eqref{temp} for $r > 0$ and $r<0$ as
\begin{eqnarray}\label{beta-plus1}
\beta_n^{+} =\frac{1}{E_n-E_0}\frac{\partial S^{+}_{a,b,r}}{\partial p_n}\qquad\text{and}\qquad
\beta_n^{-} =\frac{1}{E_n-E_0}\frac{\partial S^{-}_{a,b,r}}{\partial p_n}\,,
\end{eqnarray}
respectively.
%
%
Taking the limits $p_0 \rightarrow 1$ and $p_n \rightarrow 0$, we obtain the inverse temperature expression for $r > 0$ as 
\begin{equation} \label{beta-plus1 lim1}
\lim_{\substack{p_{0}\rightarrow1 \\ p_n\rightarrow 0} } \beta_{n}^{+}  \rightarrow \frac{2\gamma+r(\gamma-a)}{(E_n-E_0)(\gamma-a)\gamma}
\end{equation}
and likewise for $r < 0$
\begin{equation} \label{beta-minus1 lim2}
\lim_{\substack{p_{0}\rightarrow1 \\ p_n\rightarrow 0} } \beta_{n}^{-} \rightarrow \frac{2\gamma-r(\gamma-a)}{(E_n-E_0)(\gamma-a)\gamma}.
\end{equation}
The inspection of Eq. \eqref{beta-plus1 lim1} shows that in order for the group-theoretical entropy to satisfy the third law of thermodynamics for $r > 0$ i.e., $\lim_{\left\lbrace p_{0},p_{n}\right\rbrace  \to \left\lbrace 1,0\right\rbrace }  \beta_{n}^{+}\rightarrow\infty$, one has two options: the first option is that one has $\gamma-a\rightarrow 0^{+}$, which in turn  implies $\gamma= a \geq 0$, since we already have $\gamma=\sqrt{a^2+4b}\geqslant0$. The second option is that $\gamma\rightarrow0^{+}$ so that we have $\gamma=0$. However, one should also keep in mind that the third law of thermodynamics is an if and only if statement so that one should also investigate these results in a converse manner \cite{Bento}. Therefore, one asks whether the conditions $\gamma=a\geq0$ or $\gamma=0$ imply $\lim_{\left\lbrace p_{0},p_{n}\right\rbrace  \to \left\lbrace 1,0\right\rbrace } \beta_{n}^{+} \rightarrow \infty$, thereby warranting the validity of the third law of thermodynamics. To answer this question, we set $\gamma=0$ in Eq. \eqref{beta-plus1} and see that it yields $ \beta_n^{+} = 0$ independent of the values of the probabilities $\left\lbrace p_{0},p_{n}\right\rbrace $ instead of the required third law limit $\beta_{n}^{+}\rightarrow\infty$. Therefore, we are left with the option $\gamma= a > 0$ for further check. Assuming $\gamma=a$, we have
\begin{eqnarray}\label{beta-plus1 gama=a}
\nonumber
\beta_n^{+}&=&\frac{1}{(E_n-E_0)}\left(\frac{-4a p_n^{2r}-4a(r-1)p_n^{r}}{4 a^2 p_n^{2r}}+\frac{4a p_0^{2r}+4a(r-1)p_0^{r}}{4a^2 p_0^{r}}\right)\\
\nonumber
&=&\frac{1}{(E_n-E_0)a}\left(\frac{-p_n^{2r}-(r-1)p_n^{r}}{ p_n^{2r}}+\frac{ p_0^{2r}+(r-1)p_0^{r}}{ p_0^{r}}\right)\\
\nonumber
&=&\frac{1}{(E_n-E_0)a}\left(\frac{(1-r)p_n^{r}p_0^{2r}-(1-r)p_0^{r}p_n^{2r}}{p_0^{2r}p_n^{2r}}\right)\\
&=&\frac{1-r}{(E_n-E_0)a}\left(\frac{1}{p_n^{r}}-\frac{1}{p_0^{r}}\right).
\end{eqnarray}
The inspection of Eq. \eqref{beta-plus1 gama=a} indicates that the condition $\gamma=a$ together with $r \in (0,1)$ ensures the third law of thermodynamics  i.e. $\lim_{\{p_{n},p_{0}\}\to \{0,1\}} \beta_{n}^{+}\rightarrow\infty$. As a result, we reach the following conclusion for the group-theoretical entropy $S_{a,b,r}^{+}$:    
\begin{eqnarray}\label{beta-plus1 Result1}
\lim_{\substack{p_{0}\to 1\\p_{n}\to 0}} \beta_{n}^{+}\rightarrow\infty \qquad \text{if and only if} \qquad \gamma=a>0\quad \text{and} \quad 0<r<1.
\end{eqnarray}
Adopting $S_{a,b,r}^{-}$, a similar analysis yields
\begin{eqnarray}\label{beta-minus1 Result1}
\lim_{\substack{p_{0}\to 1\\p_{n}\to 0}} \beta_{n}^{-}\rightarrow\infty \qquad \text{if and only if} \qquad \gamma=a>0\quad \text{and} \quad -1<r< 0.
\end{eqnarray}
Since $\gamma=\sqrt{a^2+4b}\geqslant0$, we finally obtain for $S^+_{a,b,r}$
\begin{eqnarray}\label{new beta-plus}
\lim_{\substack{p_{0}\to 1\\p_{n}\to 0}} \beta_{n}^{+}\rightarrow\infty \qquad \qquad \text{if and only if} \qquad 0<r<1 \quad \text{and} \quad a>0\quad , \quad b=0
\end{eqnarray}
and for $S^-_{a,b,r}$
\begin{eqnarray}\label{new beta-minus}
\lim_{\substack{p_{0}\to 1\\p_{n}\to 0}} \beta_{n}^{-}\rightarrow\infty \qquad \text{if and only if} \qquad -1<r<0 \quad \text{and} \quad a>0\quad , \quad b=0
\end{eqnarray}
in terms of the genuine parameters $\{a, b, r\} $ of the group-theoretical entropy.

At this point, one might be tempted to conclude that the validity of the third law of thermodynamics is incompatible with the concavity, since there is no common interval of validity found between Eq. \eqref{new beta-plus} and Eq. \eqref{plus-condition1} for $r > 0$ (similarly, compare Eq. \eqref{new beta-minus} with Eq. \eqref{minus-condition1} for $r < 0$). However, neither Eq. \eqref{plus-condition1} nor Eq. \eqref{minus-condition1} include the case $b = 0$ and therefore this particular case should be addressed as a separate one: to proceed, we first set $b = 0$ and assume $a > 0$ to be consistent with the bounds of the third law of thermodynamics given in Eqs. \eqref{new beta-plus} and \eqref{new beta-minus}. Then, the group theoretical entropies $S_{a,r}^{+}$ for $r > 0$ and $S_{a,r}^{-}$ for $r < 0$ read 
\begin{equation}\label{Sp_with_bzero}
S_{a,r}^{+} =\frac{1}{a}
\sum_{\lambda} p_\lambda (p_{\lambda}^{-r}- 1)
\end{equation}
and 
\begin{equation}\label{Sm_with_bzero}
S_{a,r}^{-} =\frac{1}{a} \sum_{\lambda}  p_\lambda (p_{\lambda}^{r}- 1)\,,
\end{equation}
respectively. It is worth noting that these two-parameter entropies have first been obtained in Refs. \cite{new1,new2} through the group-theoretical approach while we instead reached these entropies through the third law of thermodynamics. To check the concavity, one should calculate the first and second partial derivatives with respect to $p_n$ and see whether the former is zero and the latter is negative. The first partial derivative of $S_{a,r}^{+}$ yields
\begin{eqnarray}\label{frac11}
\frac{\partial S_{a,r}^{+}}{\partial p_n} = \frac{(1-r)}{a}\left(\frac{1}{p_n^{r}}-\frac{1}{p_0^{r}}\right) = 0 \,.
\end{eqnarray}
From the above equation, it can be seen that the equality is satisfied by the uniform distribution $p_n = p_0 =p_{eq}$ if and only if $\left( r \neq 0, 1 \right) $ and $a$ is bounded. The second partial derivative with this uniform distribution yields
\begin{equation}\label{frac111}
\frac{\partial^{2} S_{a,r}^{+}}{\partial p_n^{2}} = \frac{2r(r-1)(p_{eq})^{r+1}}{a}, 
\end{equation}
which shows that the concavity is satisfied for the group-theoretical entropy  $S_{a,r}^{+}$ if the following condition is satisfied
\begin{equation}\label{new axioms1}
0 <r < 1\quad and \quad a >0\quad, \quad b=0.
\end{equation}
A similar analysis for $S_{a,r}^{-}$ shows that the concavity holds if
\begin{eqnarray}\label{new axioms2}
-1 < r < 0\quad and \quad a >0\quad, \quad b=0.
\end{eqnarray}

The two equations above show that the group-theoretical entropy takes its maximum value for the uniform distribution in the given intervals. Moreover, the comparison of the concavity conditions in Eqs. \eqref{new axioms1}-\eqref{new axioms2} with the interval of the validity of the third law of thermodynamics in Eqs. \eqref{new beta-plus}-\eqref{new beta-minus} shows that the group-theoretical entropy is both concave and conforms to the third law of the thermodynamics for $0 < r < 1$ and $-1 < r < 0$ when $a > 0$ and the parameter $b$ is set to zero.
Concerning the continuity axiom of Khinchin, the group-theoretical entropy is easily seen to be continuous with respect to all its arguments even when $a > 0$ and $b = 0$ in general \cite{tempesta2}.
The last axiom of the three is to check whether the expansibility holds.
This axiom states that the entropy does not change if a zero probability event is added. To see that it is indeed so for the group-theoretical entropy with $b = 0$ and $a > 0$, we use Eq. \eqref{Sp_with_bzero} for $0 < r < 1$ to calculate 
\begin{equation}\label{limit1}
\lim_{p \rightarrow 0} \frac{p (p^{-r}- 1)}{a} = 0
\end{equation}
and likewise Eq. \eqref{Sm_with_bzero} for $-1 < r < 0$ so that we have
\begin{equation}\label{limit2}
\lim_{p \rightarrow 0} \frac{p (p^{r}- 1)}{a} = 0.
\end{equation}

\begin{figure}
\centering
\includegraphics[width=10cm]{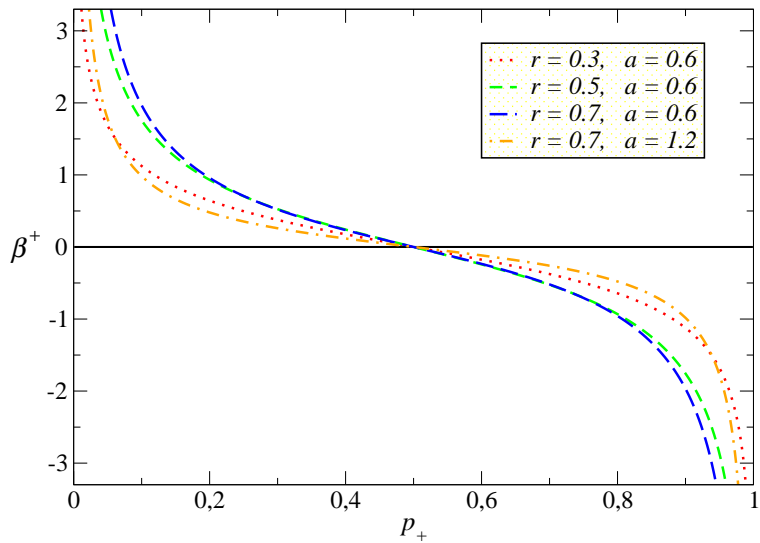}
\caption{The inverse temperature $\beta^{+}$ of the group-theoretical entropy $S_{a,r}^{+}$ in Eq. \eqref{Sp_with_bzero} versus $p_{+}$ for the one-dimensional Ising model.}
\label{Fig1}
\end{figure}

\begin{figure}
\centering
\includegraphics[width=10cm]{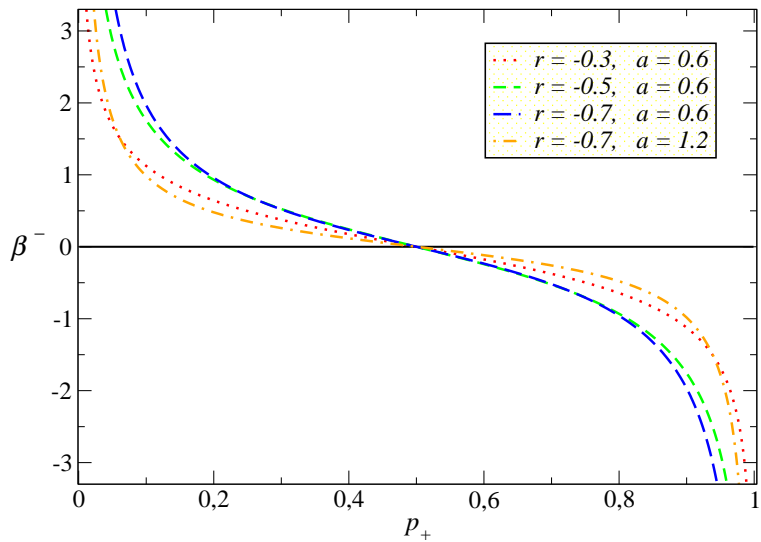}
\caption{The inverse temperature $\beta^{-}$ of the group-theoretical entropy $S_{a,r}^{-}$ in Eq. \eqref{Sm_with_bzero} versus $p_{+}$ for the one-dimensional Ising model.}
\label{Fig2}
\end{figure}

The third law of thermodynamics is expected to hold independent of the underlying Hamiltonian structure \cite{Bento}. Therefore, to illustrate our results, we consider one dimensional Ising model with periodic boundary conditions, namely, $\sigma_i=\sigma_{i+1}$, with no external field 
\begin{equation}\label{hamiltonian}
H=-J \sum_{i}\sigma_i\sigma_{i+1}\,,
\end{equation} 
where $J$ is the interaction strength and the summation is taken over the number of spins whose values are $\pm 1$. The mean energy is given by $U = J \left( p_{+} -p_{-} \right)$ where $p_{+}$ ($p_{-}$) denotes the probability of the random pair of neighboring spins being anti-parallel (parallel). Note that the mean energy can also be written as $U = J p_{+} -J \left( 1-p_{+} \right)$ due to the normalization. The energies of the anti-parallel and parallel cases are $E_1 =+J$ and $E_0 =-J$, respectively. In terms of the notation adopted before, the ground state probability $p_{0}$ corresponds to $p_{-}$ while the excited state $p_{1}$ (with $n = 1$) corresponds to $p_{+}$. Then, $\frac{\partial U}{\partial p_{n}}$ i.e. $\frac{\partial U}{\partial p_{+}}$ in Eq. \eqref{temp} is equal to $2J$ for the one-dimensional Ising model. One can then calculate the entropy using only the probability $p_{+}$, since $p_{+}=1-p_{-}$ due to the normalization. Setting $2J = 1$ without loss of generality, we plot the group-theoretical inverse temperatures for $r > 0$ and $r < 0$ in Figs. \ref{Fig1} and \ref{Fig2}, respectively. 

From these figures, it can be seen that the inverse temperature $\beta$ attains the limit $+ \infty$ when $p_{+} \rightarrow 0$ (i.e. when $p_{-} \rightarrow 1$ implying that only the ground state is populated) both for $r > 0 $ and $r < 0$. Moreover, just like the Boltzmann-Gibbs entropy, the group-theoretical inverse temperature attains the limit $- \infty$ as $p_{+} \rightarrow 1$ (i.e. when $p_{-} \rightarrow 0$ implying that only the ground state is populated) for $r > 0 $ as well as $r < 0$ as can be seen from Figs. \ref{Fig1} and \ref{Fig2}.

We finally consider the universality class of the group-theoretical entropy with respect to the classification proposed by Hanel and Thurner \cite{hanel}. In doing so, we choose the appropriate interval of parameters given by Eqs. (15)-(16) and (21)-(22) so that the group-theoretical entropy then conforms both to the third law of thermodynamics and the three axioms of Khinchin, respectively. In accordance with the formalism and notation of Ref. \cite{hanel}, we begin by calculating

\begin{eqnarray}
f(z)&=&\lim_{x\to0}\frac{g(zx)}{g(x)}
\end{eqnarray}

\noindent where $g(x)=x\frac{x^{-r}-1}{a}$ from Eq. (17) for the interval $0<r<1$. Then, one finds that $f(z) = z^{1-r}$. Comparing this with $f(z) = z^{c}$ in Ref. \cite{hanel}, one sees that $c = 1-r$. Next, we evaluate

\begin{eqnarray}
h_c(a')&=\lim_{x\to0}\frac{g(x^{1+a'})}{x^{a'c}g(x)}
\end{eqnarray}

\noindent with $a' \neq 0$ and obtain $h_c(a') = 1$. The comparison of $h_c(a') = 1$ with $h_c(a') = \left( 1 + a' \right)^{d}$ \cite{hanel} yields $d = 0$.  These results show that the universality class of the group-theoretical entropy for $0<r<1$ is determined by the scaling exponents $(c,d) = (1-r,0)$ in the large size limit.

For the interval $-1<r<0$, we have $g(x)=x\frac{x^{r}-1}{a}$ from Eq. (18). We again use Eqs. (26) and (27) and obtain the scaling exponents $(c,d) = (1+r,0)$ for $-1 < r < 0$ in the large size limit.

The asymptotic analysis above has important implications for the group-theoretical entropy: the three-parameter group-theoretical entropy $S_{a, b, r }$ for both $r > 0$ and $r < 0$ belongs to the same universality class of the Tsallis entropy $S_q$ for $q > 1$ with $(c,d) = (1,0)$ \cite{tempesta3, hanel}. However, the two-parameter group-theoretical entropy $S_{a, r }$ we obtained in this work, simultaneously conforming to the third law of thermodynamics and the three Khinchin axioms, belongs to the universality class $(c,d) = (1-r,0)$ for $0 < r < 1$ and $(c,d) = (1+r,0)$ for $-1 < r < 0$. Note that both of these universality classes are now different than that of the Tsallis entropy \cite{hanel}. In fact, the two-parameter group-theoretical entropy $S_{a, r }$ for $0 < r < 1$ is in the same universality class as the Kaniadakis entropy when one identifies $r$ as the deformation parameter $\kappa$ of Kaniadakis entropy (see Table 1 in Ref. \cite{hanel}). In the interval $-1 < r < 0$, the entropy $S_{a, r }$ describes a novel universality class.

\section{Conclusions}

Numerous new entropy measures can be found in the literature such as Tsallis \cite{Tsallis}, R{\'e}nyi \cite{Renyi}, Kaniadakis \cite{Kaniadakis} and Borges-Roditi entropies \cite{Roditi} to cite a few. Despite this multitude of entropies however, an underlying concept to classify them is difficult to obtain. One such attempt, for example, has been provided by Hanel and Thurner through solely two scaling exponents \cite{hanel}.

A novel and valuable inquiry for the common origin of the trace-form entropies has recently been proposed in terms of an underlying group-theoretical structure in Refs. \cite{tempesta1, tempesta2, tempesta3}. This unifying group structure allows one to obtain a three-parameter entropy expression in terms of the independent parameters $\{ a,b,r \}$ \cite{tempesta3}. However, some restrictions should be imposed on any entropy measure if it is to be used in the context of thermodynamics. In this sense, the third law of thermodynamics is very selective in determining the interval of the validity of the parameter-dependent entropies, and therefore it serves as a check for the generalized entropies \cite{Bento,Renyi2,third}. Moreover, since the third law should be obeyed independent of the type of interaction the system undergoes, its scope is very general.

We therefore checked the domain of applicability of the group-theoretical entropy by using the third law of thermodynamics. Whether one has $r > 0$ or $r < 0$, the group-theoretical entropy is found to conform to the third law of thermodynamics only when $b$ is set to zero. In other words, the third parameter i.e. $b$ is redundant in the face of the third law of thermodynamics thereby reducing the group-theoretical entropy to a two-parameter expression in terms of $\{ a,r \}$. We have also illustrated the interval of the validity of the third law of thermodynamics by using one-dimensional Ising model with no external field.

We have further shown that this resulting two-parameter entropy conforms to the three axioms of Khinchin i.e. continuity, concavity and expansibility exactly in the intervals required solely by the third law of thermodynamics. In the terminology of Ref. \cite{hanel}, the aforementioned result implies that this two-parameter entropy is admissible in the sense that it has the potential to serve as a candidate to describe interacting physical systems.

The restriction $b = 0$ imposed by the third law is important not only in terms of reducing the parameter space of the group-theoretical entropy, but also in terms of ensuring the extensivity. As noted in Ref. \cite{tempesta3}, when $b \neq 0$, the group-theoretical entropy is monotonically increasing function of the accessible states but not extensive. Only through the condition $b = 0$, the extensivity is ensured. In this sense, one can state that the third law of thermodynamics particularly picks the interval in which the group-theoretical entropy is extensive.

Regarding the universality class of the group-theoretical entropy, one should first note that the three-parameter group-theoretical entropy $S_{a, b, r }$ is in the same universality class of the Tsallis entropy $S_q$ for $q > 1$ \cite{tempesta3, hanel}. On the other hand, the situation is drastically different for the two-parameter group-theoretical entropy $S_{a, r }$ resulting from checking both the third law of thermodynamics and the three Khinchin axioms. In fact, it belongs to the same universality class as the Kaniadakis entropy for the interval $0 < r < 1$ while it has an universality class of its own for $-1 < r < 0$ \cite{hanel}.

We finally note that the interval of validity for the third law of thermodynamics associated with the group-theoretical entropy yields the one of the Tsallis entropy as a particular case. This can be seen by identifying $a = r = 1-q$ in Eq. (15) while one has $- a = r = 1-q$ in Eq. (16) yielding exactly the same results in Ref. \cite{third} (see in particular Eqs. (19) and (21) in Ref. \cite{third}).

\end{document}